\documentclass[%
 aip,
 jmp,
 amsmath,amssymb,
 reprint,%
]{revtex4-1}

\usepackage{graphicx}% Include figure files
\usepackage{dcolumn}% Align table columns on decimal point
\usepackage{bm}% bold math
\usepackage[utf8]{inputenc}
\usepackage[T1]{fontenc}
\usepackage{mathptmx}
\usepackage{etoolbox}

\makeatletter
\def\@email#1#2{%
 \endgroup
 \patchcmd{\titleblock@produce}
  {\frontmatter@RRAPformat}
  {\frontmatter@RRAPformat{\produce@RRAP{*#1\href{mailto:#2}{#2}}}\frontmatter@RRAPformat}
  {}{}
}%
\makeatother
\begin{document}

\preprint{AIP/123-QED}

\title{Study of stationary rigidly rotating anisotropic cylindrical fluids with new exact interior solutions of GR. 3. Azimuthal pressure.} %Title of paper

% repeat the \author .. \affiliation  etc. as needed
% \email, \thanks, \homepage, \altaffiliation all apply to the current author.
% Explanatory text should go in the []'s, 
% actual e-mail address or url should go in the {}'s for \email and \homepage.
% Please use the appropriate macro for the type of information

% \affiliation command applies to all authors since the last \affiliation command. 
% The \affiliation command should follow the other information.

\author{M.-N. C\'el\'erier}
\email{marie-noelle.celerier@obspm.fr}
\affiliation{Laboratoire Univers et Th\'eories, Observatoire de Paris, Universit\'e PSL, Universit\'e Paris Cit\'e, CNRS, F-92190 Meudon, France}

% Collaboration name, if desired (requires use of superscriptaddress option in \documentclass). 
% \noaffiliation is required (may also be used with the \author command).
%\collaboration{}
%\noaffiliation

\date{\today}

\begin{abstract}
The investigation of interior spacetimes sourced by stationary cylindrical anisotropic fluids is pursued and specialized here to rigidly rotating fluids with an azimuthally directed pressure. Based on the occurence of an extra degree of freedom in the equations, two general methods for constructing different classes of exact solutions to the field equations are proposed. Exemplifying such recipes, a bunch of solutions are constructed. Axisymmetry and regularity conditions on the axis are examined and the spacetimes are properly matched to a vacuum exterior. A number of classes and subclasses are thus studied and an analysis of their features leads to sorting out three classes whose appropriate mathematical and physical properties are discussed.  This work is part of a larger study of the influence of anisotropic pressure in GR, using cylindrical symmetry as a simplifying assumption, and considering in turn each principal stress direction. It has been initiated in companion Papers 1 and 2 where the pressure was assumed to be axially directed, and is followed by Paper 4 considering radial pressure and Paper 5 contrasting the previous results with corresponding dust and perfect fluid solutions.
\end{abstract}

% insert suggested PACS numbers in braces on next line

\maketitle %\maketitle must follow title, authors, abstract and \pacs

% Body of paper goes here. Use proper sectioning commands. 
\section{Introduction} \label{intro}

\onecolumngrid % print on one column

This article is the third in a series of five where the study of interior solutions to the field equations of General Relativity (GR) sourced by a cylinder of stationary fluid with anisotropic pressure rigidly rotating around its symmetry axis have been reported. For mathematical purposes, the problem has been divided into three distinct cases each corresponding to a special direction of the pressure. In Papers 1 \cite{C21} and 2 \cite{C22a} it was axially directed and its azimuthal and radial components were assumed to vanish. In the present article, which is denoted Paper 3, the pressure is azimuthal which implies that its axial and radial components vanish. Accordingly, in the subsequent Paper 4, the pressure is radially directed since its two other components are zero. Paper 5 is devoted to a discussion about the previous results, including a comparison with analogous dust and perfect fluid solutions.

In Paper 1, a particular exact solution to the axial pressure issue has been displayed. Paper 2 is a generalization of this preliminary result. Actually, owing to the presence of an extra degree of freedom in the calculations, this case is likely to give rise to a number of different solutions, each corresponding to a given physical configuration. Hence, a general method allowing one to generate these different classes of solutions has been constructed and is described in that article where some particular classes have been integrated and analyzed.

Two other methods, based on the extra degree of freedom also available here, have been developed to construct azimuthal pressure solutions. They are displayed in the present Paper 3 where they are also exemplified through three different classes each corresponding to some particular physical configuration.

A proposal of possible applications of such kind of results has already been given in Paper 2 to which the interested reader is referred. Note that, for a better overview of the issue, reading the full set of five papers is recommended.

The outlook of the present article reads as follows. In Sec. \ref{ci}, the general equations determining the interior spacetimes, i.e., the field equations and the Bianchi identity, are displayed. Section \ref{gm} is devoted to the description of a general method for constructing exact solutions with azimuthal pressure. It is a key result of this work, since it allows one to built at will new solutions other than those depicted here. Indeed, in Sec. \ref{ex}, this method is applied to two different classes of spacetimes for which thorough mathematical and physical analyses are conducted. In Sec. \ref{versus}, the main features of these classes of solutions are compared. Another method, adapted to the special form of the field equations corresponding to the present case of azimuthally directed pressure, is displayed in Sec. \ref{other} where the solutions for fluids with the widely used equation of state $P_{\phi} = h \rho$, $h=const.$, are derived and discussed. This method is another key result of this work, since it allows one to choose a priori a given equation of state for the fluid, provided of course that the corresponding equations are integratable. The conclusions are given in Sec. \ref{concl}.

\section{Equations determining the interior spacetime} \label{ci}

The cylinder of fluid is stationary and rigidly rotating around its axis of symmetry. It is an anisotropic nondissipative fluid bounded by a cylindrical surface $\Sigma$. Its principal stresses $P_r$, $P_z$ and $P_\phi$ satisfy the equation of state $P_r= P_z=0$, which allows one to write its stress-energy tensor -- see (1) of C\'el\'erier and Santos \cite{CS20} for its general expression -- under the form
\begin{equation}
T_{\alpha \beta} = \rho V_\alpha V_\beta + P_{\phi} K_\alpha K_\beta, \label{setens}
\end{equation}
with $\rho$, the energy density of the fluid, $V_\alpha$, its timelike 4-velocity, and $K_\alpha$, a spacelike 4-vector, verifying
\begin{equation}
V^\alpha V_\alpha = -1, \quad  K^\alpha K_\alpha = 1, \quad V^\alpha K_\alpha = 0. \label{fourvec}
\end{equation}
The spacelike Killing vector $\partial_z$ is hypersurface orthogonal, such as to ease its subsequent matching to an exterior Lewis metric. The line element reads therefore
\begin{equation}
\textrm{d}s^2=-f \textrm{d}t^2 + 2 k \textrm{d}t \textrm{d}\phi +\textrm{e}^\mu (\textrm{d}r^2 +\textrm{d}z^2) + l \textrm{d}\phi^2, \label{metric}
\end{equation}
with $f$, $k$, $\mu$, and $l$, real functions of the radial coordinate $r$ only, such as to allow for stationarity. Due to cylindrical symmetry, the coordinates are bond to conform to the following ranges:
\begin{equation}
-\infty \leq t \leq +\infty, \quad 0 \leq r \leq +\infty, \quad -\infty \leq z \leq +\infty, \quad 0 \leq \phi \leq 2 \pi, \label{ranges}
\end{equation}
with the two limits of the coordinate $\phi$ topologically identified. These coordinates are denoted $x^0=t$, $x^1=r$, $x^2=z$, and $x^3=\phi$.

In the case of rigid rotation, a frame corotating with the fluid can be chosen \cite{CS20,C21, D06}. Thus, the 4-velocity of the fluid can be written as
\begin{equation}
V^\alpha = v \delta^\alpha_0, \label{r4velocity}
\end{equation}
with $v$ a function of $r$ only. Therefore, the timelike condition for $V^\alpha$ displayed in (\ref{fourvec}) reads
\begin{equation}
fv^2 = 1. \label{timelike}
\end{equation}
The spacelike 4-vector $K^{\alpha}$ satisfying conditions (\ref{fourvec}) can be written as
\begin{equation}
K^\alpha = - \frac{kv}{D}\delta^\alpha_0 - \frac{fv}{D}\delta^\alpha_3, \label{kalpha}
\end{equation}
where the auxiliary function $D(r)$ is still defined as
\begin{equation}
D^2 = fl + k^2. \label{D2}
\end{equation}

\subsection{Field equations} \label{fe}

Using (\ref{r4velocity})--(\ref{D2}) into (\ref{setens}), the components of the stress-energy tensor corresponding to the five nonvanishing components of the Einstein tensor are obtained, and the following five field equations for the spacetime inside $\Sigma$ can be written as

\begin{equation}
G_{00} = \frac{\textrm{e}^{-\mu}}{2} \left[-f\mu'' - 2f\frac{D''}{D} + f'' - f'\frac{D'}{D} + \frac{3f(f'l' + k'^2)}{2D^2}\right] = \kappa\rho f, \label{G00}
\end{equation}
\begin{equation}
G_{03} =  \frac{\textrm{e}^{-\mu}}{2} \left[k\mu'' + 2 k \frac{D''}{D} -k'' + k'\frac{D'}{D} - \frac{3k(f'l' + k'^2)}{2D^2}\right] = - \kappa\rho k, \label{G03}
\end{equation}
\begin{equation} 
G_{11} = \frac{\mu' D'}{2D} + \frac{f'l' + k'^2}{4D^2} = 0, \label{G11}
\end{equation}
\begin{equation}
G_{22} = \frac{D''}{D} -\frac{\mu' D'}{2D} - \frac{f'l' + k'^2}{4D^2} = 0, \label{G22}
\end{equation}
\begin{equation}
G_{33} =  \frac{\textrm{e}^{-\mu}}{2} \left[l\mu'' + 2l\frac{D''}{D} - l'' + l'\frac{D'}{D} - \frac{3l(f'l' + k'^2)}{2D^2}\right]  =  \frac{\kappa}{f} \left(\rho k^2 + P_{\phi} D^2\right), \label{G33}
\end{equation}
where the primes stand for differentiation with respect to $r$.

\subsection{Conservation of the stress-energy tensor -- Bianchi identity} \label{bi}

The equation for the conservation of the stress-energy tensor is analogous to the Bianchi identity:
\begin{equation}
T^\beta_{1;\beta} = 0. \label{Bianchi}
\end{equation}
Writing the twice contravariant version of (\ref{setens}), one obtains
\begin{equation}
T^{\alpha \beta} = \rho V^\alpha V^\beta + P_{\phi} K^\alpha K^\beta, \label{setenscontra}
\end{equation}
where $V^\alpha$ is given by (\ref{r4velocity}), and the spacelike vector $K^\alpha$ is given by (\ref{kalpha}), which are inserted into (\ref{setenscontra}). Using (\ref{metric}) and (\ref{timelike}), the Bianchi identity (\ref{Bianchi}) reduces to
\begin{equation}
T^\beta_{1;\beta} = \frac{1}{2} \left(\rho + P_{\phi} \right) \frac{f'}{f} - P_{\phi}\frac{D'}{D} = 0. \label{Bianchi2}
\end{equation}
With $h(r)$ defined as $h(r)\equiv P_{\phi}(r)/\rho(r)$, (\ref{Bianchi2}) can be written as
\begin{equation}
\frac{1}{2} (1+h) \frac{f'}{f} - h \frac{D'}{D} = 0. \label{Bianchi3}
\end{equation}

\section{General method for constructing the solutions} \label{gm}

Adding both field equations (\ref{G11}) and (\ref{G22}) gives
\begin{equation}
D'' = 0, \label{sol1}
\end{equation}
which can be integrated as
\begin{equation}
D = c_1r+c_2 \label{sol2}
\end{equation}
where $c_1$ and $c_2$ are integration constants.

The coordinate $r$ can then be rescaled from a factor $c_1$, which gives
\begin{equation}
D = r+c_2. \label{sol3}
\end{equation}

Here again, as in the axial pressure case, only five independent differential equations are available for six unknowns, i.e., the four functions $f$, $k$, $\textrm{e}^\mu$, and $l$, the energy density $\rho$, and the pressure defined either by $P_{\phi}$ or by $h$. Therefore, the set of equations needs to be closed by an additional one. Assuming the choice to be that of a given expression $f(h)$ for the $f$ metric function, its derivative with respect to $r$ can be written as
\begin{equation}
f' = \frac{\textrm{d} f}{\textrm{d} h} h', \label{sol4}    
\end{equation}
that yields
\begin{equation}
\frac{f'}{f} = \frac{\frac{\textrm{d} f}{\textrm{d} h}}{f} h', \label{sol5}    
\end{equation}
which gives, once inserted into the Bianchi identity (\ref{Bianchi3}),
\begin{equation}
\frac{D'}{D} = \frac{(1+h)}{2h}\frac{\frac{\textrm{d} f}{\textrm{d} h}}{f} h', \label{sol6}    
\end{equation}
which can be integrated as
\begin{equation}
\ln{D} = \frac{1}{2}\int _{h_0}^h \frac{(1+u)}{u}\frac{\frac{\textrm{d} f}{\textrm{d} u}}{f} \textrm{d}u. \label{sol7}    
\end{equation}
Hence, $f(h)$ must be chosen such that the above integral should be fully integrated yielding thus an analytical expression for $D(h)$, which, differentiated with respect to $r$, gives
\begin{equation}
D' = \frac{\textrm{d} D}{\textrm{d} h} h'. \label{sol8}    
\end{equation}
Now, (\ref{sol3}) differentiated with respect to $r$, yields
\begin{equation}
D' = 1, \label{sol9}    
\end{equation}
which one equalizes to (\ref{sol8}) and obtains
\begin{equation}
h' = \frac{1}{\frac{\textrm{d} D}{\textrm{d} h}}. \label{sol10}  \end{equation}

Now, the same reasoning as in Papers 1 and 2 combining (\ref{G00}) with (\ref{G03}), then integrating, gives
\begin{equation}
kf' - fk' = 2c D, \label{sol11}
\end{equation}
where $2c$ is an integration constant and where the factor 2 is chosen for further convenience. Considered as a first-order ordinary differential equation for $k(r)$, (\ref{sol11}) possesses as a general solution
\begin{equation}
k = f \left(c_k - 2c\int_{h_0}^h \frac{D(v)}{f(v)^2 h'(v)} \textrm{d}v \right), \label{sol12}
\end{equation}
where $c_k$ is a new integration constant and $h_0$ is the value of $h$ on the axis of symmetry. With (\ref{sol10}) inserted, (\ref{sol12}) becomes
\begin{equation}
k = f \left(c_k - 2c\int_{h_0}^h \frac{D(v)}{f(v)^2} \frac{\textrm{d} D}{\textrm{d}v}\textrm{d}v \right). \label{sol13}
\end{equation}
Provided the above integral is solvable, an analytical expression for $k(h)$ is thus obtained. Then, $l$ can be calculated using (\ref{D2}) as
\begin{equation}
l = \frac{D^2 - k^2}{f}. \label{sol14}    
\end{equation}

Now, using (\ref{D2}) into (\ref{sol11}), one obtains the expression for $(f'l'+k'^2)/2D^2$ derived in Papers 1 and 2 and recalled here as
\begin{equation}
\frac{f'l'+k'^2}{2D^2} = \frac{f'D'}{fD} - \frac{f'^2}{2f^2} + \frac{2c^2}{f^2}, \label{sol14b}
\end{equation}
which, inserted into (\ref{G11}), yields
\begin{equation}
\frac{\mu'D'}{D} + \frac{f'D'}{fD} - \frac{f'^2}{2f^2} + \frac{2c^2}{f^2} =0, \label{sol15}
\end{equation}
where the Bianchi identity (\ref{Bianchi3}) is substituted, that gives
\begin{equation}
\frac{1+h}{2h}\left(\frac{\frac{\textrm{d} f}{\textrm{d} h}}{f}\right) h' \mu' + \frac{1}{2h}\left(\frac{\frac{\textrm{d} f}{\textrm{d} h}}{f}\right)^2 h'^2 + \frac{2c^2}{f^2} = 0. \label{sol16}
\end{equation}
Multiplying the last term of (\ref{sol16}) by $h'^2/h'^2 = h'^2 (\textrm{d} D/\textrm{d} h)^2 = 1$, one obtains
\begin{equation}
\frac{1+h}{2h}\left(\frac{\frac{\textrm{d} f}{\textrm{d} h}}{f}\right) h' \mu' + \frac{1}{2h}\left(\frac{\frac{\textrm{d} f}{\textrm{d} h}}{f}\right)^2 h'^2 + \frac{2c^2}{f^2} \left(\frac{\textrm{d} D}{\textrm{d} h}\right)^2 h'^2 = 0, \label{sol17}
\end{equation}
that gives
\begin{equation}
\mu' = -\frac{2h}{1+h}\left(\frac{f}{\frac{\textrm{d} f}{\textrm{d} h}}\right)\left[\frac{1}{2h}\left(\frac{\frac{\textrm{d} f}{\textrm{d} h}}{f}\right)^2 + \frac{2c^2}{f^2} \left(\frac{\textrm{d} D}{\textrm{d} h}\right)^2 \right] h'. \label{sol18}
\end{equation}
Provided the right-hand-side is integrable, an expression for $\mu(h)$ follows.

Then, the axisymmetry condition \cite{S09,P96,M93}
\begin{equation}
l\stackrel{0}{=}0 \label{axic}
\end{equation}
and, if appropriate, the regularity condition \cite{D06,S09,M93}
\begin{equation}
\frac{\textrm{e}^{-\mu} l'^2}{4l} \stackrel{0}{=} 1, \label{regc}
\end{equation}
where $\stackrel{0}{=}$ denotes the value of the functions taken on the rotation axis, i.e., for $r=0$, are implemented and give constraints on the integration constant parameters .

To enable applications of the resulting spacetimes to a representation of astrophysical processes, a proper junction to an exterior vacuum metric is necessary. As it has been chosen and justified in Papers 1 and 2, the Weyl class of the Lewis solution \cite{L32} is selected to represent the vacuum exterior. Darmois' junction conditions \cite{D27} are thus applied to both inside and outside metrics on the $\Sigma$ boundary. As mentioned in Papers 1 and 2, these conditions come down to $P_r = 0$ on the boundary \cite{I66}. Since, by virtue of the equation of state imposed in the present configuration, $P_r = 0$ everywhere, this condition is obviously verified in particular on the $\Sigma$ hypersurface.

Now, to derive an expression for the energy density $\rho(h)$ (\ref{G00}) is arranged with the help of (\ref{G11}) and (\ref{sol1}) to become
\begin{equation}
-\mu'' + \frac{f''}{f} - \frac{f'D'}{fD} - \frac{3 \mu' D'}{D}= 2\kappa \textrm{e}^{\mu}\rho, \label{rho1}   
\end{equation}
from which $\rho$ is extracted, since, from the preceding, $f(h)$, $D(h)$, $\mu(h)$ and $h'(h)$ are known.

Then, $P_{\phi}$ follows, from the definition of $h$, as
\begin{equation}
P_{\phi} = h \rho. \label{1cp}
\end{equation}

Once the metric and the physical properties issued from the field equations coupled to the choice of $f(h)$ are determined, other different mathematical and physical features can be exhibited and analyzed as it has been done in Papers 1 and 2 for the equations of state studied there.

\section{Exemplifying the general method with fully integrated solutions} \label{ex}

To exemplify the method displayed in Section \ref{gm}, two classes of exact solutions are proposed here. They are obtained through two different expressions for the metric function $f(h)$ and their properties are analyzed and contrasted.

\subsection{Class 1 solutions}

The expression for the metric function $f$ is chosen to be
\begin{equation}
f= \frac{c_f}{1+h}. \label{1c1}    
\end{equation}
Differentiated with respect to $r$, this becomes
\begin{equation}
f'= - c_f \frac{h'}{(1+h)^2}. \label{1c2}    
\end{equation}
Then, (\ref{1c2}) divided by (\ref{1c1}) gives
\begin{equation}
\frac{f'}{f}= - \frac{h'}{1+h}, \label{1c3}    
\end{equation}
which, substituted into (\ref{Bianchi3}), yields
\begin{equation}
\frac{D'}{D}= - \frac{h'}{2h}, \label{1c4}    
\end{equation}
that can be integrated as
\begin{equation}
D= \frac{1}{c_1 \sqrt{h}}, \label{1c5}    
\end{equation}
which can be inserted into (\ref{sol3}) and give
\begin{equation}
\frac{1}{\sqrt{h}}= c_1 r + c_1 c_2. \label{1c6}
\end{equation}
The coordinate $r$ can thus be rescaled from a factor $c_1 \rightarrow 1$ which yields
\begin{equation}
\frac{1}{\sqrt{h}}= r + c_2. \label{1c7}    
\end{equation}
At the axis, where $r=0$, $h$ is denoted $h_0$ and (\ref{1c7}) gives
\begin{equation}
c_2 = \frac{1}{\sqrt{h_0}}, \label{1c8}    
\end{equation}
which, inserted into (\ref{1c7}), yields $h(r)$ as
\begin{equation}
h = \frac{1}{\left( r+\frac{1}{\sqrt{h_0}}\right)^2}, \label{1c9} \end{equation}
which implies, when differentiated with respect to $r$,
\begin{equation}
h' = - \frac{2}{\left( r+\frac{1}{\sqrt{h_0}}\right)^3}, \label{1c10} 
\end{equation}
where one can insert (\ref{1c9}) such as to obtain
\begin{equation}
h' = - 2 h \sqrt{h}. \label{1c11} 
\end{equation}
Some straightforward relations, which will be needed further on, follow as
\begin{equation}
D= \frac{1}{\sqrt{h}}, \label{1c12}    
\end{equation}
\begin{equation}
D= r + \frac{1}{\sqrt{h_0}}, \label{1c13}    
\end{equation}
\begin{equation}
r = \frac{1}{\sqrt{h}} - \frac{1}{\sqrt{h_0}}. \label{1c14}    
\end{equation}

Now, (\ref{1c1}), (\ref{1c11}) and (\ref{1c12})  are inserted into (\ref{sol12}) that gives
\begin{equation}
k = \frac{c_f}{1+h} \left(c_k + \frac{c}{c_f^2}\int_{h_0}^h \frac{(1+v)^2}{v^2} \textrm{d}v \right), \label{1c15}
\end{equation}
which can be integrated as
\begin{equation}
k = \frac{c_f}{1+h} \left[c_k + \frac{c}{c_f^2} \left( \frac{1}{h_0} - \frac{1}{h} + 2 \ln \frac{h}{h_0} + h - h_0 \right) \right]. \label{1c16}
\end{equation}
Then $l$ follows from (\ref{sol14}) as
\begin{equation}
l = \frac{1+h}{c_f h} -\frac{c_f}{1+h} \left[c_k + \frac{c}{c_f^2} \left( \frac{1}{h_0} - \frac{1}{h} + 2 \ln \frac{h}{h_0} + h - h_0 \right) \right]^2. \label{1c17}
\end{equation}

To calculate $\textrm{e}^{\mu}$, $f$, $D$ and derivatives are inserted into (\ref{sol15}) which becomes
\begin{equation}
\mu' = \frac{h'}{(1+h)^2} + \frac{c^2}{c_f^2} \frac{(1+h)^2 h'}{h^2}, \label{1c18}
\end{equation}
which can be integrated as
\begin{equation}
\mu = - \frac{1}{1+h} + \frac{c^2}{c_f^2} \left(-\frac{1}{h} + 2\ln h + h \right) + \ln c_{\mu}, \label{1c19}
\end{equation}
where $c_{\mu}$ is an integration constant. Equation (\ref{1c19}) can be written as
\begin{equation}
\textrm{e}^{\mu} = c_{\mu} h^{\frac{2c^2}{c_f^2}} \exp \left[ - \frac{1}{1+h} + \frac{c^2}{c_f^2} \left(h-\frac{1}{h} \right) \right]. \label{1c20}
\end{equation}
Now, the integration constant $c_{\mu}$ can be set to $1$ by rescaling both coordinates $r$ and $z$ from a factor $\sqrt{c_{\mu}}$. Hence, $c_{\mu}$ disappears from the expression (\ref{1c20}) of $\textrm{e}^{\mu}$.

\subsubsection{Axisymmetry and regularity conditions} \label{1caxi}

For this class of solutions, the axisymmetry condition (\ref{axic}) becomes
\begin{equation}
c_f c_k = \frac{1 + h_0}{\sqrt{h_0}}, \label{1cax}    
\end{equation}
where the sign indeterminacy issued from taking the square root has been absorbed into the definition of the constants.

The regularity condition given by (\ref{regc}) reads in this case
\begin{equation}
c_f= 2c \frac{(1 + h_0)^2}{\sqrt{h_0}(h_0 - 1)}, \label{1creg}    
\end{equation}
which, substituted into (\ref{1cax}), yields
\begin{equation}
c_k = \frac{h_0-1}{2c(1+h_0)}. \label{1caxreg}    
\end{equation}

\subsubsection{Updated expressions of the metric functions}

Inserting (\ref{1creg}) and (\ref{1caxreg}) into (\ref{1c1}), (\ref{1c16}), (\ref{1c17}) and (\ref{1c20}), one obtains new expressions for the metric functions that read
\begin{equation}
f= \frac{2c (1+h_0)^2}{\sqrt{h_0}(h_0 - 1)(1+h)}, \label{1c21}    
\end{equation}
\begin{equation}
k =  \frac{1+h_0}{\sqrt{h_0}(1+h)} \left[1 + \frac{h_0(h_0 - 1)}{2(1 + h_0)^3} \left( \frac{1}{h_0} - \frac{1}{h} + 2 \ln \frac{h}{h_0} + h - h_0 \right) \right], \label{1c22}
\end{equation}
\begin{equation}
l = \frac{\sqrt{h_0}(h_0 - 1)}{2c (1+h_0)^2}\frac{(1+h)}{h} - \frac{(h_0 -1)}{2c \sqrt{h_0} (1+h)} \left[1 + \frac{h_0(h_0 - 1)}{2(1+h_0)^3} \left( \frac{1}{h_0} - \frac{1}{h} + 2 \ln \frac{h}{h_0} + h - h_0 \right) \right]^2, \label{1c23}
\end{equation}
\begin{equation}
\textrm{e}^{\mu} = h^{\frac{h_0(1-h_0)^2}{2(1+h_0)^4}} \exp \left[ - \frac{1}{1+h} + \frac{h_0(1-h_0)^2}{4(1+h_0)^4}\left(h-\frac{1}{h} \right) \right]. \label{1c24}
\end{equation}
At this stage, the metric depends only on two parameters, $h_0$ and $c$. However, it will be seen in Section \ref{c1hyd} that $c$ can be given an interpretation allowing it to be written as an expression involving only $h_0$ which will thus remain the single parameter characterizing each metric in class 1.

\subsubsection{Behaviour of the $h(r)$ function} \label{1chr}

The evolution of the ratio $h$ from the axis to the $\Sigma$ boundary is now considered. At the axis, a new constraint on its value $h_0$ can be exhibited from considering (\ref{1creg}). Actually, to obtain a proper signature for the metric, the four metric functions have to be either all positive or all negative definite. Now, the occurrences of $\sqrt{h}$ and of $\sqrt{h_0}$ into (\ref{1c5})--(\ref{1c14}) imply $h$ and $h_0$ positive since, for astrophysical purpose, the functions and parameters of the solutions are imposed to be real valued. From (\ref{1c24}), the positiveness of $h$ implies that of $\textrm{e}^{\mu}$. Therefore, all the metric functions have to be positive definite which is, in particular, the case for $f$. Hence, from (\ref{1c1}), the parameter $c_f$ must be positive. From (\ref{1creg}), $c_f$ positive requires $c$ and $h_0-1$ both positive or both negative. The first case implies $h_0>1$ while the second one compels $h_0<1$. Since, from the foregoing, $h_0$ is positive, both cases are actually satisfied, and the second one reduces to $0<h_0<1$. The metric function $f$ is therefore positive definite as required.

Now, since the $r$ coordinate is anywhere positive or null, the expression (\ref{1c10}) of the first derivative of $h(r)$ with respect to $r$ shows that $h'<0$ whatever $r$. Hence, the $h(r)$ function is monotonically decreasing from $h_0>0$ to $h_{\Sigma} > 0$ on the boundary.

\subsubsection{The energy density and pressure} \label{c1edp}

To obtain $\rho(h)$, $f$ given by (\ref{1c21}), $\textrm{e}^{\mu}$ given by (\ref{1c24}), $D$ given by (\ref{1c12}) and their derivatives, where $h'$ given by (\ref{1c11}) is substituted, are inserted into (\ref{rho1}) such as to give
\begin{equation}
\rho = \frac{4}{\kappa h^{\frac{h_0(1-h_0)^2}{2(1+h_0)^4}}}\left[\frac{h_0(1-h_0)^2}{4(1+h_0)^4}(1+h) - \frac{h^2}{(1+h)^3}\right]
\exp \left[ \frac{1}{1+h} + \frac{h_0(1-h_0)^2}{4(1+h_0)^4}\left(\frac{1}{h} - h \right) \right]. \label{1c25}
\end{equation}

Now, the weak energy condition, $\rho \geq 0$, implies
\begin{equation}
w = \frac{h_0(1-h_0)^2}{4(1+h_0)^4}(1+h) - \frac{h^2}{(1+h)^3} \geq 0. \label{1c26}
\end{equation}
On the axis where $h=h_0$, inequality (\ref{1c26}) becomes
\begin{equation}
1-6h_0 + h_0^2 \geq 0, \label{1c27}
\end{equation}
which imposes
\begin{equation}
h_0>3+2\sqrt{2} \qquad \textrm{or} \qquad h_0<3-2\sqrt{2}. \qquad \label{1c28}
\end{equation}

The result obtained in Section \ref{1chr}, i.e., $h_0>0$, is consistent with both constraints in (\ref{1c28}), i.e., either it reads $h_0>3+2\sqrt{2}$ or it reduces to $0<h_0<3-2\sqrt{2}$.

Assuming one or the other above inequality is satisfied, if $w$ does not vanish before reaching the boundary $\Sigma$, it remains always positive and the weak energy condition is fulfilled in the whole interior spacetime.

However, if $w$ vanishes at some $r$ value, denoted $r_0$, smaller than $r_{\Sigma}$, there will be a cylindrical shell where $\rho<0$, implying that the weak energy condition is not satisfied. Now, $w$ vanishes if
\begin{equation}
\frac{h^2}{(1+h)^4} = \frac{h_0(1-h_0)^2}{4(1+h_0)^4}, \label{1c26a}
\end{equation}
of which we take the square root that yields a second degree equation in $h$ which can be written as
\begin{equation}
\frac{\sqrt{h_0}(1-h_0)}{2(1+h_0)^2} + \left[\frac{\sqrt{h_0}(1-h_0)}{(1+h_0)^2} -1\right]h + \frac{\sqrt{h_0}(1-h_0)}{2(1+h_0)^2}h^2 =0, \label{1c26b}
\end{equation}
which possesses real roots, provided the discriminant $\Delta$ is positive or null. Now, since $\Delta$ reads
\begin{equation}
\Delta = 1 - \frac{2\sqrt{h_0}(1-h_0)}{(1+h_0)^2}, \label{1c26c}
\end{equation}
it is easy to infer that, for both admissible intervals for $h_0$, $\Delta>0$ and $w$ vanishes for two real values of $h$, collectively denoted $h_a$ and expressed as
\begin{equation}
h_a = \frac{(1+h_0)^2}{\sqrt{h_0}(1-h_0)} \left(1 \pm \sqrt{\Delta} \right) -1. \label{1c26d}
\end{equation}

Therefore, for each value of $h_0$ defining a particular solution belonging to Class 1, there exists a limiting value of $r_{\Sigma}$, denoted $r_a$, and such that
\begin{equation}
r_a = \frac{1}{\sqrt{h_a}} - \frac{1}{\sqrt{h_0}}, \label{1c26e}
\end{equation}
owing to (\ref{1c14}).

Now, the energy density as a function of the radial coordinate $r$ is obtained by substituting (\ref{1c9}) into (\ref{1c25}) which gives
\begin{eqnarray}
&&\frac{\kappa \rho}{4} = \left(r+\frac{1}{\sqrt{h_0}}\right)^{\frac{h_0(1-h_0)^2}{(1+h_0)^4}} \left\{\frac{h_0(1-h_0)^2}{4(1+h_0)^4}\left[1+ \frac{1}{\left(r+\frac{1}{\sqrt{h_0}}\right)^2}\right]
- \frac{1}{\left(r+\frac{1}{\sqrt{h_0}}\right)^4 \left[1+\frac{1}{\left(r+\frac{1}{\sqrt{h_0}}\right)^2}\right]^3}\right\} \nonumber \\
&\times& \exp \left\{ \frac{1}{1+\frac{1}{\left(r+\frac{1}{\sqrt{h_0}}\right)^2}}
+\frac{h_0(1-h_0)^2}{4(1+h_0)^4}\left[\left(r+\frac{1}{\sqrt{h_0}}\right)^2 - \frac{1}{\left(r+\frac{1}{\sqrt{h_0}}\right)^2} \right] \right\}. \label{1c30a}
\end{eqnarray}

An analysis of the function $\rho(r)$, not detailed here since it is rather lengthy and cumbersome, gives the following interesting result. The energy density $\rho$ decreases from $r=0$ to $r_{\Sigma}$, the coordinate of the boundary having indeed to satisfy $r_{\Sigma} \leq r_a$. Considering that the pressure $P_{\phi}$ can be written as $P_{\phi} = h \rho$ and that both the functions $\rho(r)$ and $h(r)$ are decreasing when $r$ increases, it comes that $P_{\phi}(r)$ is also a decreasing function of $r$ which never vanishes inside the cylinder.

Moreover, from (\ref{1c9}), it is obvious that $h$ is positive whatever the value of $r$ which conforms the statement in Sec. \ref{1chr}. Hence the weak energy condition $\rho>0$ implies the strong energy condition, $P_{\phi}>0$. 

\subsubsection{Metric signature and sign constraints} \label{c1ms}

So that Class 1 solutions be eligible, the metric must exhibit a Lorentzian signature which implies, due to the positiveness of $\textrm{e}^{\mu}$, that all the other metric functions should be positive definite.

Owing to $c_f >0$ and $h>0$, the metric function $f$ given by (\ref{1c1}) is indeed correct from this point of view.

Now, $h_0$ and $h$ being positive, the sign of $k$ is given by the second factor in (\ref{1c22}). This second factor is obviously positive near the axis where $h \sim h_0$. However, for each given value of $h_0$, there exists a ratio $h_1$, defined by
\begin{equation}
1 + \frac{h_0(h_0 - 1)}{2(1 + h_0)^3} \left( \frac{1}{h_0} - \frac{1}{h_1} + 2 \ln \frac{h_1}{h_0} + h_1 - h_0 \right) =0, \label{1c30}
\end{equation}
and such that $k>0$ for $h>h_1$ and $k<0$ for $h<h_1$. To avoid an improper sign for $k$ when approaching the boundary the value of $h$ there must verify $h_{\Sigma}>h_1$.

The sign of $l$ is now considered. From the axisymmetry condition, at the axis, $l(h_0)=0$. The derivative of $l$ with respect to $h$ evaluated in the vicinity of the axis, i.e., for $h=h_0 - \epsilon$, with $\epsilon$ a small positive increment, reads 
\begin{equation}
\frac{\textrm{d}l}{\textrm{d}h} = - \frac{\epsilon (h_0 - 1)}{4c h_0^{\frac{5}{2}}} (1+h_0)^3\left(-1 + 5 h_0 + 2 h_0^{\frac{5}{2}} \right), \label{1c31}
\end{equation}
which is negative provided
\begin{equation}
-1 + 5 h_0 + 2 h_0^{\frac{5}{2}}>0. \label{1c31b}
\end{equation}
Recall indeed that, as required in Sec. \ref{1chr}, $(h_0-1)/c$ is positive and so is $h_0$. Moreover, inequality (\ref{1c31b}) is actually satisfied for positive values of $h_0$ such that $h_0>0.193419$. Hence, near the axis, $l$ is decreasing with $h$ increasing, itself decreasing with $r$ increasing. Therefore, $l$ is increasing with $r$ increasing. Since it has already been shown that $l'$ vanishes only for $r=0$, $l'$ cannot change sign elsewhere. Thus, $l$ is monotonically increasing from $l=0$ at the axis to $l(h_{\Sigma})$ on the boundary.

Every metric function being positive definite, the signature of the metric exhibits the proper feature, provided (\ref{1c31b}) is satisfied.

\subsubsection{Hydrodynamical properties} \label{c1hyd}

As it is well-known, the timelike 4-vector $V_{\alpha}$ can be invariantly decomposed into the acceleration vector, the rotation or twist tensor and the shear tensor. The expressions for these quantities have been displayed by C\'el\'erier and Santos \cite{CS20} for a rotating cylindrical fluid such as the one considered here.

The only nonzero component of the acceleration vector can be written as
\begin{equation}
\dot{V}_1 = \frac{1}{2} \frac{f'}{f}, \label{1c32}
\end{equation}
which becomes, with (\ref{1c3}) and (\ref{1c11}) inserted,
\begin{equation}
\dot{V}_1 = \frac{h\sqrt{h}}{1+h}. \label{1c33}
\end{equation}
Its modulus follows as
\begin{equation}
\dot{V}^{\alpha}\dot{V}_{\alpha} = \frac{1}{4} \frac{f'^2}{f^2} \textrm{e}^{-\mu}, \label{1c34}
\end{equation}
which becomes, with (\ref{1c3}) and (\ref{1c24}) inserted,
\begin{equation}
\dot{V}^{\alpha}\dot{V}_{\alpha} = \frac{h^{3-\frac{h_0(1-h_0)^2}{2(1+h_0)^4}}}{(1+h)^2} \exp\left[\frac{1}{1+h}+ \frac{h_0(1-h_0)^2}{4(1+h_0)^4}\left(\frac{1}{h} -h\right)\right]. \label{1c35}
\end{equation}
The rotation scalar takes the form
\begin{equation}
\omega^2 = \frac{c^2}{f^2\textrm{e}^{\mu}}, \label{1c36}
\end{equation}
which becomes, after inserting (\ref{1c1}) and (\ref{1c24}),
\begin{equation}
\omega^2 = \frac{h_0^{1-\frac{h_0(1-h_0)^2}{2(1+h_0)^4}}(1-h_0)^2}{4(1+h_0)^4} (1+h)^2 \exp\left[\frac{1}{1+h}+ \frac{h_0(1-h_0)^2}{4(1+h_0)^4}\left(\frac{1}{h} -h\right)\right]. \label{1c37}
\end{equation}

On the axis, its value is
\begin{equation}
\omega^2 \stackrel{0}{=} \frac{h_0^{1-\frac{h_0(1-h_0)^2}{2(1+h_0)^4}}(1-h_0)^2}{4(1+h_0)^2} \exp\left[\frac{1}{1+h_0}+ \frac{(1-h_0)^3}{4(1+h_0)^3}\right]. \label{1c38}
\end{equation}

As it has been shown in Appendix A of Paper 2, the $c$ parameter measures the amplitude of the rotation scalar, provided (19) and (A1) of Paper 2 are satisfied which is indeed the case here. Therefore, $c$ can be obtained as the square root of (\ref{1c38}) that gives
\begin{equation}
c = \frac{h_0^{\frac{1}{2}-\frac{h_0(1-h_0)^2}{4(1+h_0)^4}}(h_0-1)}{2(1+h_0)} \exp\left[\frac{1}{2(1+h_0)}+ \frac{(1-h_0)^3}{8(1+h_0)^3}\right], \label{1c39}
\end{equation}
the sign in (\ref{1c39}) being chosen such that the signs of $c$ and $h_0-1$ be equal as required in Section \ref{1chr}.

As it is well-known and already recalled in Papers 1 and 2, rigid rotation implies a vanishing shear.

\subsubsection{Parameter of the solutions}

Inserting (\ref{1c39}) into (\ref{1c21}) and (\ref{1c23}), one obtains metric coefficients depending on a single parameter which is the value of the ratio $h_0$ on the axis.

Each Class 1 solution is therefore fully determined once the ratio $h_0$ has been measured or imposed on the axis provided of course it should be larger than $0.193419$.

\subsubsection{Singularities}

An interesting property of this class is that its metrics do not exhibit any singularity. This is due to the limit imposed on the ratio $h$, which prevent the metric functions to vanish or to diverge.

\subsubsection{Final form of the class 1 solutions} \label{c1final}

In order to present in a concise form the main quantities describing the solutions of this class, the metric functions, the energy density, the pressure and the main intermediate functions are updated to their final form and summarized below.
\begin{equation}
f = \frac{1+h_0}{h_0^{\frac{h_0(1-h_0)^2}{4(1+h_0)^4}}(1+h)} \exp\left[\frac{1}{2(1+h_0)}+ \frac{(1-h_0)^3}{8(1+h_0)^3}\right], \label{1cf1}
\end{equation}
\begin{equation}
\textrm{e}^{\mu} = h^{\frac{h_0(1-h_0)^2}{2(1+h_0)^4}} \exp \left[ - \frac{1}{1+h} + \frac{h_0(1-h_0)^2}{4(1+h_0)^4}\left(h-\frac{1}{h} \right) \right], \label{1cf2}
\end{equation}
\begin{equation}
k =  \frac{1+h_0}{\sqrt{h_0}(1+h)} \left[1 + \frac{h_0(h_0 - 1)}{2(1 + h_0)^3} \left( \frac{1}{h_0} - \frac{1}{h} + 2 \ln \frac{h}{h_0} + h - h_0 \right) \right], \label{1cf3}
\end{equation}
\begin{eqnarray}
l &=& h_0^{\frac{h_0(1-h_0)^2}{4(1+h_0)^4}} \exp\left[-\frac{1}{2(1+h_0)} - \frac{(1-h_0)^3}{8(1+h_0)^3}\right] \left\{\frac{(1+h)}{(1+h_0)h} - \frac{1+h_0}{h_0(1+h)} \right.
\nonumber \\
&\times& \left. \left[1 + \frac{h_0(h_0 - 1)}{2(1+h_0)^3} \left( \frac{1}{h_0} - \frac{1}{h} + 2 \ln \frac{h}{h_0} + h - h_0 \right) \right]^2\right\}, \label{1cf4}
\end{eqnarray}
\begin{equation}
\rho = \frac{4}{\kappa h^{\frac{h_0(1-h_0)^2}{2(1+h_0)^4}}} \left[\frac{h_0(1-h_0)^2}{4(1+h_0)^4}(1+h) - \frac{h^2}{(1+h)^3}\right] \exp\left[\frac{1}{(1+h)} + \frac{h_0(1-h_0)^2}{4(1+h_0)^4}\left(\frac{1}{h} - h \right)\right],  \label{1cf5}
\end{equation}
\begin{equation}
P_{\phi} = \frac{4 h^{1-\frac{h_0(1-h_0)^2}{2(1+h_0)^4}}}{\kappa} \left[\frac{h_0(1-h_0)^2}{4(1+h_0)^4}(1+h) - \frac{h^2}{(1+h)^3}\right]   \exp\left[\frac{1}{(1+h)} + \frac{h_0(1-h_0)^2}{4(1+h_0)^4}\left(\frac{1}{h} - h \right)\right],  \label{1cf6}
\end{equation}
\begin{equation}
D = \frac{1}{\sqrt{h}} = r + \frac{1}{\sqrt{h_0}}, \label{1cf7}
\end{equation}
\begin{equation}
h = \frac{1}{\left(r + \frac{1}{\sqrt{h_0}}\right)^2}, \label{1cf8}
\end{equation}
\begin{equation}
h' = -2h\sqrt{h} = -\frac{2}{\left(r + \frac{1}{\sqrt{h_0}}\right)^3}, \label{1cf9}
\end{equation}
\begin{equation}
r = \frac{1}{\sqrt{h}} - \frac{1}{\sqrt{h_0}}. \label{1cf10}
\end{equation}

The function $h(r)$ evolves according to
\begin{equation}
h_0>h>h_{\Sigma}>\textrm{Max}\{h_1,0.193419\}>0, \label{1cf12}
\end{equation}
$h_1$ being defined by (\ref{1c30}).

\subsection{Class 2 solutions}

\subsubsection{Integration of the field equations} \label{2cint}

The choice of the $f(h)$ metric function made here, equivalent to the one made in Paper 1 dealing with axially directed pressures, amounts to setting
\begin{equation}
\frac{f'}{f} = \frac{2h'}{1-h}, \label{ex1}
\end{equation}
which can be integrated as
\begin{equation}
f = \frac{c_f}{(1-h)^2}, \label{ex2}
\end{equation}
where $c_f$ is an integration constant.
Inserting (\ref{ex1}) into the Bianchi identity (\ref{Bianchi3}), one obtains
\begin{equation}
\frac{1+h}{h(1-h)}h' = \frac{D'}{D}, \label{ex3}
\end{equation}
which can be written as
\begin{equation}
\frac{h'}{h}+\frac{2h'}{1-h} = \frac{D'}{D}, \label{ex4}
\end{equation}
and then, integrated as
\begin{equation}
D=  \frac{h}{c_5(1-h)^2}, \label{ex5}
\end{equation}
$c_5$ being another integration constant.

Now, (\ref{sol3}) inserted into (\ref{ex5}) yields
\begin{equation}
\frac{h}{(1-h)^2} = c_5r + c_2 c_5. \label{ex6}
\end{equation}
Thus, the coordinate $r$ can, once again, be rescaled, now from a factor $c_5$, and then (\ref{ex6}) becomes
\begin{equation}
\frac{h}{(1-h)^2} = r + c_2, \label{ex7}
\end{equation}
with $c_5=1$.

Differentiating (\ref{ex7}) with respect to $r$ yields
\begin{equation}
h'= \frac{(1-h)^3}{1+h}, \label{ex8}
\end{equation}
which is substituted, together with (\ref{ex2}) and (\ref{ex5}), into (\ref{sol12}) such as to obtain
\begin{equation}
k = \frac{c_f}{(1-h)^2} \left[c_k - \frac{2c}{c_f^2}\int_{h_0}^h \frac{v(1+v)}{1-v}  \textrm{d}v \right]. \label{ex9}
\end{equation} 
This equation can be integrated as
\begin{equation}
k = \frac{c_f}{(1-h)^2} \left\{c_k - \frac{2c}{c_f^2} \left[2 \ln \left(\frac{1-h_0}{1-h}\right) + 2(h_0-h) + \frac{h_0^2 - h^2}{2}\right] \right\}. \label{ex10}
\end{equation}
Thus, the metric function $l$ follows from (\ref{D2}) as
\begin{equation}
l = \frac{h^2}{c_f(1-h)^2} - \frac{c_f}{(1-h)^2}\left\{c_k-  \frac{2c}{c_f^2}\left[2 \ln \left(\frac{1-h_0}{1-h}\right) + 2(h_0-h) + \frac{h_0^2 - h^2}{2} \right] \right\}^2. \label{ex11}
\end{equation}

Then, (\ref{ex1}) and (\ref{ex2}) inserted into (\ref{sol17}) give
\begin{equation}
\frac{(1+h)h'}{h(1-h)}\mu' + \frac{2h'^2}{h(1-h)^2} + \frac{2c^2}{c_f^2}\frac{(1+h)^2 h'^2}{(1-h)^2} = 0, \label{ex12}
\end{equation}
that can  be written as
\begin{equation}
\mu' = - \frac{h'}{(1-h)} - \frac{h'}{(1+h)} + \frac{2c^2}{c_f^2}\left[hh' + 2h' - \frac{2 h'}{(1-h)}\right], \label{ex13}
\end{equation}
which can be integrated as
\begin{equation}
\mu = \ln\frac{(1-h)}{(1+h)} + \frac{4c^2}{c_f^2}\ln(1-h) + \frac{c^2}{c_f^2}(h^2+ 4h)  + \ln c_{\mu}, \label{ex14}
\end{equation}
where $c_{\mu}$ is another integration constant, and that yields
\begin{equation}
\textrm{e}^{\mu} = c_{\mu}\frac{(1-h)^{1+\frac{4c^2}{c_f^2}}}{(1+h)} \exp\left[\frac{c^2}{c_f^2}h(4+h)\right].  \label{ex15}
\end{equation}

\subsubsection{Energy density and pressure} \label{2cedp}

To obtain an expression for $\rho$, the two metric functions $f$ and $\textrm{e}^{\mu}$, together with the intermediate function $D$, and their derivatives, are inserted into (\ref{rho1}) such as to yield
\begin{equation}
\rho = \frac{2(1-h)^{3-\frac{4c^2}{c_f^2}}}{\kappa c_{\mu}} \left[ \frac{1-h}{h(1+h)^3} + \frac{2c^2}{c_f^2}\right] \exp\left[- \frac{c^2}{c_f^2}h(4+h)\right]. \label{2crho}
\end{equation}

The pressure $P_{\phi}$ follows by multiplying (\ref{2crho}) by $h$ which gives
\begin{equation}
P_{\phi} = \frac{2(1-h)^{4-\frac{4c^2}{c_f^2}}}{\kappa c_{\mu}} \left[ \frac{1-h}{h(1+h)^3} + \frac{2c^2}{c_f^2}\right] \exp\left[- \frac{c^2}{c_f^2}h(4+h)\right]. \label{2cp}
\end{equation}

\subsubsection{Axisymmetry and regularity conditions} \label{regcond}

To represent an axisymmetric spacetime the solution obtained above must verify (\ref{axic}). With (\ref{ex11}) inserted into this condition, it yields
\begin{equation}
l \stackrel{0}{=} \frac{h_0^2}{c_f (1-h_0)^2} - \frac{c_f c_k^2}{(1-h_0)^2} = 0, \label{regcond1}
\end{equation}
which can be written
\begin{equation}
c_f c_k = h_0, \label{regcond2}
\end{equation}
where the sign indeterminacy due to the extraction of the square root has been absorbed into the definition of the integration constants.

Now, the regularity condition (\ref{regc}) with the first derivative of (\ref{ex11}) and with (\ref{ex15}) inserted, yields
\begin{equation}
(1+h_0) \left[\frac{2h_0}{c_f(1+h_0)} - \frac{2c_f c_k^2}{1+h_0} + \frac{4c c_k h_0}{c_f}\right] =0, \label{regcond3}
\end{equation}
which has two solutions. First
\begin{equation}
h_0 = -1. \label{regcond4}
\end{equation}
This value of $h_0$ substituted into (\ref{regcond2}) gives
\begin{equation}
c_k = -\frac{1}{c_f}. \label{regcond5}
\end{equation}
The second solution becomes, after inserting (\ref{regcond2}) into the second factor of (\ref{regcond3}),
\begin{equation}
c_f = -\frac{2c h_0 (1+h_0)}{1-h_0}, \label{regcond6}
\end{equation}
which substituted into (\ref{regcond2}) gives
\begin{equation}
c_k = -\frac{(1 - h_0)}{2c (1+h_0)}. \label{regcond7}
\end{equation}

Now, evaluating both expressions of $D$, (\ref{sol3}) and (\ref{ex5}), at the axis where $r=0$ and $h=h_0$, while using $c_5=1$, one obtains
\begin{equation}
c_2 = \frac{h_0}{(1-h_0)^2}. \label{i1}
\end{equation}

\subsubsection{Subclass (i): generalities} \label{sci}

It is the subclass of class 2 which includes the solutions verifying (\ref{regcond4}), hence (\ref{regcond5}). Inserting (\ref{regcond4}) into (\ref{i1}) yields
\begin{equation}
c_2 = -\frac{1}{4}, \label{i2}
\end{equation}
which gives, once inserted into (\ref{sol3}),
\begin{equation}
D = r - \frac{1}{4}, \label{i3}
\end{equation}
and once inserted into (\ref{ex7}),
\begin{equation}
r = \frac{h}{(1-h)^2} + \frac{1}{4}. \label{i4}
\end{equation}

Now, inserting (\ref{regcond4}) and (\ref{regcond5}) into (\ref{ex10}), then into (\ref{ex11}), give
\begin{equation}
k = - \frac{1}{(1-h)^2} \left\{1 + \frac{2c}{c_f} \left[2 \ln \left(\frac{2}{1-h}\right) - 2(1+h) + \frac{1-h^2}{2}\right] \right\}, \label{i5}
\end{equation}
\begin{equation}
l = \frac{1}{c_f(1-h)^2} \left\lgroup h^2 - \left\{1 + \frac{2c}{c_f}\left[2 \ln \left(\frac{2}{1-h}\right) - 2(1+h) + \frac{1-h^2}{2} \right] \right\}^2 \right\rgroup. \label{i6}
\end{equation}

In this case, the expressions (\ref{2crho}) and (\ref{2cp}) for the energy density and the azimuthal pressure stay unchanged.

\subsubsection{Subclass (ii): generalities} \label{scii}

It is the subclass of class 2 which includes the solutions verifying (\ref{regcond6}) and (\ref{regcond7}). Here, there is no definite value imposed to $h_0$, therefore $c_2$ is merely given by (\ref{i1}), which yields, once inserted into (\ref{sol3}),
\begin{equation}
D = r + \frac{h_0}{(1-h_0)^2}, \label{ii1}
\end{equation}
and once inserted into (\ref{ex7}),
\begin{equation}
r = \frac{h}{(1-h)^2} - \frac{h_0}{(1-h_0)^2}. \label{ii2}
\end{equation}

Now, inserting (\ref{regcond6}) and (\ref{regcond7}) into (\ref{ex2}), (\ref{ex10}), (\ref{ex11}) and (\ref{ex15}) gives the expressions for the metric functions of this subcase (ii) as
\begin{equation}
f = \frac{2c h_0(1+h_0)}{(h_0-1)(1-h)^2}, \label{ii2a}
\end{equation}
\begin{equation}
k = \frac{h_0}{(1-h)^2} \left\{1 + \frac{1-h_0}{h_0^2(1+h_0)} \left[2 \ln \left(\frac{1-h_0}{1-h}\right) + 2(h_0-h) + \frac{h_0^2 - h^2}{2}\right] \right\}, \label{ii3}
\end{equation}
\begin{equation}
l = \frac{h_0(1-h_0)}{2c (1+h_0)(1-h)^2} \left\lgroup \left\{1 + \frac{1-h_0}{h_0^2(1+h_0)} \left[2 \ln \left(\frac{1-h_0}{1-h}\right) + 2(h_0-h) + \frac{h_0^2 - h^2}{2} \right] \right\}^2 - \frac{h^2}{h_0^2}\right\rgroup, \label{ii4}
\end{equation}
\begin{equation}
\textrm{e}^{\mu} = c_{\mu} \frac{(1-h)^{1+\frac{(1-h_0)^2}{h_0^2(1+h_0)^2}}}{(1+h)} \exp\left[\frac{(1-h_0)^2}{4h_0^2(1+h_0)^2}h(4+h)\right].  \label{ii5}
\end{equation}

Then, inserting (\ref{regcond6}) and (\ref{regcond7}) into (\ref{2crho}) and (\ref{2cp}), one obtains the expressions for the energy density and the azimuthal pressure for case (ii) as
\begin{equation}
\rho = \frac{2(1-h)^{3 - \frac{(1-h_0)^2}{h_0^2(1+h_0)^2}}}{\kappa c_{\mu}} \left[ \frac{1-h}{h(1+h)^3} + \frac{(1-h_0)^2}{2h_0^2(1+h_0)^2}\right] \exp\left[- \frac{(1-h_0)^2}{4 h_0^2(1+h_0)^2}h(4+h)\right]. \label{2crhoii}
\end{equation}

Then, $P_{\phi}$ follows by multiplying (\ref{2crhoii}) by $h$.

\subsubsection{Behaviour of the $h(r)$ function} \label{2ch}

Equalizing both expressions of $D$, (\ref{sol3}) and (\ref{ex5}) with $c_5 = 1$, one obtains $h$ as a function of the radial coordinate $r$ that reads
\begin{equation}
h=\frac{1 +2(r+c_2) +\epsilon \sqrt{1 + 4 (r + c_2)}}{2(r+c_2)}, \label{2chr1}
\end{equation}
where $\epsilon = \pm 1$. Then, differentiating (\ref{2chr1}) with respect to $r$, one obtains
\begin{equation}
h'= - \frac{\epsilon(1+3r+3c_2) + \sqrt{1+4(r+c_2)}}{(r+c_2)^2 \sqrt{1 + 4 (r + c_2)}}. \label{2chr2}
\end{equation}
The analysis of the $h(r)$ behaviour is now straightforward.

First, notice that the expression inside the square roots must be positive to obtain real-valued $h$ and $h'$. The extremum of this expression being reached for $r=0$, this implies 
\begin{equation}
c_2 \geq - \frac{1}{4}. \label{2chr3}
\end{equation}
The equality is achieved within subcase (i), owing to (\ref{i2}), while, for subcase (ii), it is easy to verify that the inequality is satisfied whatever the value of $h_0$, and, of course, whatever the sign of $\epsilon$. 

Another feature independent of the sign of $\epsilon$ proceeds from (\ref{ex7}) where the consequences of $r>0$ are considered.
The ratio $h$ and thus the pressure can only vanish provided $c_2 \leq 0$. For subclass (i), $c_2 = -1/4$ and the condition is fulfilled without imposing extra constraints. For subclass (ii), (\ref{i1}) implies that the pressure can only vanish if, at the axis $h_0<0$.

One can also conclude from (\ref{ex8}) that, whatever $\epsilon$,
\begin{equation}
h'>0  \quad \textrm{if} \quad -1<h<+1, \label{2chr4}
\end{equation}
\begin{equation}
h'<0  \quad \textrm{if} \quad h<-1 \quad \textrm{or} \quad h>+1. \label{2chr5}
\end{equation}

Now, the sign of $\epsilon$ drives other aspects of the behaviour of $h(r)$ which are described below. 

\paragraph{Subclass (i)}

This subclass obeys (\ref{regcond4}) and (\ref{i2}) which, once inserted into (\ref{2chr1}) and (\ref{2chr2}), give
\begin{equation}
h=\frac{1 + 4r + 4 \epsilon \sqrt{r}}{4r -1}, \label{2chr6}
\end{equation}
\begin{equation}
h'= - \frac{\epsilon\left(\frac{1}{4}+3r\right) + 2\sqrt{r}}{2\left(r-\frac{1}{4}\right)^2 \sqrt{r}}, \label{2chr7a}
\end{equation}
From (\ref{2chr7a}), a possible vanishing of $h'$ is seen to occur for a negative value of the radial coordinate $r$ which is never realized. Therefore $h'$ is bound to keep the same sign in the whole spacetime and the function $h(r)$ is monotonically increasing or decreasing depending on the sign of $h'$. Still from (\ref{2chr7a}) this sign appears to be the inverse of the sign of $\epsilon$, therefore: $h(r)$ should be monotonically increasing for $\epsilon =-1$ and monotonically decreasing for $\epsilon =+1$.

\subparagraph{Case $\epsilon=+1$.}

From (\ref{2chr6}), it is easy to see that, in this case, $h$ diverges towards $-\infty$ in the vicinity of $r=1/4$, which would imply a diverging negative pressure at this locus. This drawback can be prevented provided the radial coordinate of the boundary $r_\Sigma$ satisfies $r_\Sigma < 1/4$. In this case, $h$ is decreasing from $h_0=-1$ at the axis to $h_{\Sigma}$ on the boundary and is diverging if ever $r_\Sigma \rightarrow 1/4$.

\subparagraph{Case $\epsilon=-1$.}

A more standard configuration is obtained for $\epsilon =-1$. In this case $h(r)$ is monotonically increasing with $r$. From (\ref{2chr6}), it is easy to see that, in the vicinity of $r=1/4$, $h \rightarrow 1$. Moreover, (\ref{2chr6}) still yields $h=-1$ at the axis, i.e., $h_0=-1$. The behaviour of $h$ in this case can thus be summarized as such:

- for $r_{\Sigma} <1/4$, $h$ is increasing from $h_0=-1$ at the axis to $h_{\Sigma}<0$. Hence the strong energy condition is nowhere satisfied.

- for $r_{\Sigma} >1/4$, $h$ is increasing from $h_0=-1$ at the axis to some $h_{\Sigma}>0$. Hence the strong energy condition, $h>0$, is only satisfied inside some cylindrical shell near the boundary of the spacetime. Such a behaviour lacking any known physical meaning, this case is very unlikely.

\paragraph{Subclass (ii)}

The $c_2$ parameter of this subclass is determined by (\ref{i1}) with no a priori constraint on the value of $h_0$.

\subparagraph{Case $\epsilon = +1$.}

Inserting (\ref{i1}) and the sign of $\epsilon$ into (\ref{2chr1}), it can be inferred that $h$ vanishes at a locus $r$ given by
\begin{equation}
- \frac{2h_0}{(1-h_0)^2} = 1+2r+ \sqrt{1 + 4 \left[r + \frac{h_0}{(1-h_0)^2}\right]}, \label{2chr10}
\end{equation}
which implies $h_0<0$.

Now, for the case $h_0>0$, (\ref{2chr1}) shows that $h>0$ all along from $r=0$ to $r_{\Sigma}$ and (\ref{2chr2}) implies $h'<0$ for any $r$.
Hence, the ratio $h(r)$ decreases monotonically from $h_0>0$ to $h_{\Sigma}>0$.

From (\ref{2chr5}), $h'<0$ and $h>0$ would imply $h>+1$ and therefore $1<h_{\Sigma}<h<h_0$. However, an analysis of $\rho(h)$ given by (\ref{2crhoii}) shows that, for any couple $\{h_0,h\}$ such that $h_0>1$ and $h>1$, $\rho<0$ and therefore the weak energy condition is not satisfied which implies the ruling out of the case $\epsilon = +1$ in Subclass (ii).

\subparagraph{Case $\epsilon = -1$.}

Inserting $\epsilon = -1$ and (\ref{i1}) into (\ref{2chr2}), one can conclude that $h'$ vanishes for two values of $r$ which are
\begin{equation}
r_1= - \frac{h_0}{(1-h_0)^2} , \label{2chr7}
\end{equation}
for which $r>0$ imposes $h_0<0$. And
\begin{equation}
r_2= - \frac{2}{9}- \frac{h_0}{(1-h_0)^2} , \label{2chr8}
\end{equation}
which is positive provided $-2<h_0<-1/2$. Hence, whenever the pressure is positive at the axis, $h'$ never vanishes in this subclass with $\epsilon=-1$ and the $h(r)$ function is monotonically increasing or decreasing depending on the sign of $h'$. Moreover, for $h_0>0$, the inequality
\begin{equation}
\sqrt{1+4r+ \frac{4h_0}{(1-h_0)^2}} < 1+ 2r + \frac{2h_0}{(1-h_0)^2} < 1 + 3r + \frac{3h_0}{(1-h_0)^2} \label{2chr9}
\end{equation}
holds and gives, when compared to (\ref{2chr2}), $h'> 0$. Now, from (\ref{2chr4}), $h'>0$ implies $h<+1$ and, in particular, $h_{\Sigma}<+1$.

A summary of the above remarks for subclass (ii) and case $\epsilon = -1$ runs as follows. The behaviour of the ratio $h$ depends on its initial value $h_0$ at the axis. For $h_0>0$, $h(r)$ is monotonically increasing from $h_0$ at $r=0$ to $h_{\Sigma}$ verifying $0<h_{\Sigma}<+1$ at the boundary. For $-1/2<h_0<0$ or $h_0<-2$, $h'$ vanishes once at $r=- h_0/(1-h_0)^2$. For $-2<h_0<-1/2$, $h'$ vanishes twice, at $r=- h_0/(1-h_0)^2$ and $r= -2/9 - h_0/(1-h_0)^2$.

\subsubsection{Metric signature and sign constraints} \label{2cms}

\subparagraph{Subclass (i)}

To display a proper Lorentzian signature $+2$ or equivalently $-2$, the metric functions must be all positive or all negative definite in the whole spacetime. It is in particular sufficient to show that this property does not hold in the vicinity of the axis to rule out the whole subclass. Its defining property $h_0= -1$ will be needed to carry out the reasoning. In the vicinity of $h_0$, the ratio $h$ can be written as
\begin{equation}
h_a = -1 + \eta, \label{2chr10a}
\end{equation}
where $\eta$ is a small positive increment.

The metric function $f(h_a)$, as given by (\ref{ex2}), thus reads
\begin{equation}
f(h_a) = \frac{c_f}{(2 - \eta)^2}, \label{2chr10b}
\end{equation}
which has the same sign as $c_f$.

The metric function $\textrm{e}^{\mu}$, given by (\ref{ex15}), becomes, in the vicinity of the axis,
\begin{equation}
\textrm{e}^{\mu} = c_{\mu} \frac{2^{1+\frac{4c^2}{c_f^2}}}{\eta} \exp\left(\frac{-3c^2}{c_f^2}\right).  \label{2chr10c}
\end{equation}
which has the same sign as $c_{\mu}$.

As regards $k$, described by (\ref{i5}), its behaviour near the axis is dominated by
\begin{equation}
k(h_0) = -\frac{1}{4}. \label{2chr10d}
\end{equation}
The function $k$ is therefore negative definite.

Finally, the metric function $l$, given by (\ref{i6}), becomes near $r=0$
\begin{equation}
l(h_a) = -\frac{\eta}{4c_f}(2-\eta), \label{2chr10e}
\end{equation}
which has an inverse sign with respect to that of $c_f$.

Both metric functions $f$ and $l$ exhibiting an inverse sign in a given region of the class (i) spacetimes, the signature is improper and class (i) is therefore ruled out.

\subparagraph{Subclass (ii)}

For this subclass, the value of $h_0$ is not fixed a priori as it was for subclass (i). Therefore, the reasoning will have to take into account a wide range of $h_0$ values that will be restricted in the course of the analysis. Here, the behaviours of the metric functions are more involved and more entangled. Therefore, they have been subjected to a numerical analysis as follows.

For a number of values of $h_0$ in the allowed interval $-1<h_0<+1$, the functions $k(h)$ and $2c l(h)$, given by (\ref{ii3}) and (\ref{ii4}), have been plotted. Then, for each couple of plots, one has identified the common range of $h$, including $h_0$ for continuity sake, for which each function $k$ and $cl$ are either positive or negative definite, independently one from the other. Indeed, the actual sign of $l$ has then been adjusted to that of $k$, by virtue of a choice of the sign of $c$. Now, given the sign of $c$, (\ref{ii2a}) allows to specify that of $f$ in an interval including $h_0$. If $f$, $k$ and $l$ happen to be all three positive or all three negative on some range of $h$, the sign of $\textrm{e}^{\mu}$ is adjusted by a choice of the sign of $c_{\mu}$ and the allowed range of $h$ is once more reduced if it happens to include unity where $\textrm{e}^{\mu}/c_{\mu}$ can be shown to change sign.

Actually, from (\ref{regcond6}) inserted into (\ref{ex15}), one obtains
\begin{equation}
\frac{\textrm{e}^{\mu}}{c_{\mu}} =\frac{(1-h)^{1+\frac{(1-h_0)^2}{h_0^2(1+h_0)^2}}}{(1+h)} \exp\left[\frac{(1-h_0)^2}{4h_0^2(1+h_0)^2}h(4+h)\right],  \label{2cm1}
\end{equation}
which yields, whatever the value of $h_0$,

- $\textrm{e}^{\mu}/c_{\mu} >0 \Leftrightarrow h<1 \Rightarrow h_{\Sigma}<1$

- $\textrm{e}^{\mu}/c_{\mu} <0 \Leftrightarrow h>1$ and $1+ (1-h_0)^2/h_0^2(1+h_0)^2 \neq 2n$, with $n$ integer. However, it will be shown in the following that $h>1$ never occurs in a well-behaved spacetime of this class.

Once this analysis completed for a particular value of $h_0$, either one obtains that for a given range of $h$ -- bounded by $h_{\Sigma}$ -- the four metric functions exhibit the same sign and therefore the corresponding $h_0$ is identified as defining a well-behaved spacetime. Note that if this sign is positive, the signature is $+2$, if it is negative, the signature is $-2$. However, if no common range of common sign including $h_0$ can be found, the corresponding value of $h_0$ is ruled out as not able to determine a well-behaved spacetime within this class.

Then, another value of $h_0$ undergoes an analogous treatment, and so on until a range of proper $h_0$ values is identified. For this class 2 (ii) of solutions, the allowed range is
\begin{equation}
-1<h_0<-0.42. \label{2cm2}
\end{equation}
Each value of $h_0$ between these two limits determines an allowed interval for $h$ reading
\begin{equation}
-1<h_{\Sigma}<h<h_0<0. \label{2cm3}
\end{equation}
For every genuine solution of this class, the signature of the metric is $-2$ and both signs of the parameters $c$ and $c_{\mu}$ are negative. 

\subsubsection{Hydrodynamical properties} \label{2chyd}

The hydrodynamical properties of subclass (ii) fluids are displayed below, following the scheme developed in Section \ref{c1hyd}.

The only nonzero component of the acceleration vector is obtained by inserting (\ref{ex1}) and (\ref{ex8}) into (\ref{1c32}) such as to obtain
\begin{equation}
\dot{V}_1 = \frac{(1-h)^2}{1+h}. \label{2ch1}
\end{equation}
The modulus of this vector follows as
\begin{equation}
\dot{V}^{\alpha}\dot{V}_{\alpha} = \frac{(1-h)^{3-\frac{(1-h_0)^2}{h_0^2(1+h_0)^2}}}{1+h} \exp\left[- \frac{(1-h_0)^2}{4h_0^2(1+h_0)^2}h(4+h)\right]. \label{2ch2}
\end{equation}
The rotation scalar emerges, after inserting (\ref{ii2a}) and (\ref{ii5}) into (\ref{1c36}), from
\begin{equation}
\omega^2 = \frac{(h_0-1)^2}{4h_0^2(1+h_0)^2}\frac{(1-h)^{3-\frac{(1-h_0)^2}{h_0^2(1+h_0)^2}}}{1+h}  \exp\left[- \frac{(1-h_0)^2}{4h_0^2(1+h_0)^2}h(4+h)\right], \label{2ch3}
\end{equation}

which becomes, once evaluated on the axis,
\begin{equation}
\omega^2 \stackrel{0}{=} \frac{(h_0-1)^{5-\frac{(1-h_0)^2}{h_0^2(1+h_0)^2}}}{4h_0^2(1+h_0)^3} \exp\left[- \frac{(1-h_0)^2}{4h_0^2(1+h_0)^2}h_0(4+h_0)\right]. \label{2ch4}
\end{equation}

As it has been shown in Appendix A of Paper 2, the rotation scalar is equal to the squared value of the $c$ parameter. The following expression for $c$ emerges therefore from (\ref{2ch4}):
\begin{equation}
c = -\frac{(h_0-1)^{\frac{5}{2}-\frac{(1-h_0)^2}{2h_0^2(1+h_0)^2}}}{2h_0(1+h_0)^{\frac{3}{2}}} \exp\left[- \frac{(1-h_0)^2}{8h_0(1+h_0)^2}(4+h_0)\right], \label{2ch5}
\end{equation}
where the minus sign in (\ref{2ch5}) has been chosen so that $c$ be always negative when $h_0$ spans its whole allowed range (\ref{2cm3}).

As already recalled in Section \ref{c1hyd}, rigid rotation implies a vanishing shear.

\subsubsection{Parameter of the solutions}

Owing to the form of the metric function $\textrm{e}^{\mu}$ as given by (\ref{2cm1}) the coordinate $r$ and $z$ can be rescaled from a factor $\sqrt{-c_{\mu}}$ which implies $c_{\mu}=-1$ in (\ref{2cm1}).

Inserting (\ref{2ch5}) into (\ref{ii2a}) and (\ref{ii4}), one obtains metric coefficients depending on a single parameter which is the value $h_0$ of the ratio $h$ on the axis. Each Class 2 (ii) solution is therefore fully determined once the ratio $h_0$ has been measured on the axis or imposed, provided of course (\ref{2cm2}) should be satisfied.

\subsubsection{Weak energy condition}

Now, the weak energy condition $\rho \geq 0$ is considered. Inserting $c_{\mu}=-1$ into (\ref{2crho}), one obtains
\begin{equation}
\rho = -\frac{2(1-h)^{3-\frac{4c^2}{c_f^2}}}{\kappa} \left[ \frac{1-h}{h(1+h)^3} + \frac{2c^2}{c_f^2}\right] \exp\left[- \frac{c^2}{c_f^2}h(4+h)\right]. \label{2cwe1}
\end{equation}
Since $-1<h<0$, the weak energy condition reduces to
\begin{equation}
\frac{1-h}{h(1+h)^3} + \frac{2c^2}{c_f^2} \leq 0, \label{2cwe2}
\end{equation}
which can be written
\begin{equation}
0 \leq \frac{2c^2}{c_f^2} \leq \frac{h-1}{h(1+h)^3}. \label{2cwe3}
\end{equation}
With (\ref{regcond6}) inserted, (\ref{2cwe3}) becomes
\begin{equation}
0 \leq \frac{(1-h_0)^2}{2h_0^2(1+h_0)^2} \leq \frac{h-1}{h(1+h)^3}, \label{2cwe4}
\end{equation}
which defines, for each particular solution, i.e., for each value of $h_0$, an admissible range for $h$, between $-1$ and $0$, such that (\ref{2cwe4}) is verified and the weak energy condition is fulfilled.

\subsubsection{Singularities}

For this class also, the metrics do not exhibit any singularity. This is due once more to the values imposed to the ratio $h$, which prevent the metric functions to vanish or to diverge.

\subsubsection{Final form of the Class 2 solutions}

Since it has been shown that subclass (i) is ruled out, class 2 will subsequently denote previous class 2 subclass (ii). Taking into account all the previous results and conventions which have been used to build the expressions of the functions determining the solutions, one obtain
\begin{equation}
f = -\frac{(h_0-1)^{\frac{3}{2}-\frac{(1-h_0)^2}{2h_0^2(1+h_0)^2}}}{(1+h_0)^{\frac{1}{2}}(1-h)^2} \exp\left[- \frac{(1-h_0)^2}{8h_0(1+h_0)^2}(4+h_0)\right], \label{2cf1}
\end{equation}
\begin{equation}
\textrm{e}^{\mu} =-\frac{(1-h)^{1+\frac{(1-h_0)^2}{h_0^2(1+h_0)^2}}}{(1+h)} \exp\left[\frac{(1-h_0)^2}{4h_0^2(1+h_0)^2}h(4+h)\right],  \label{2cf2}
\end{equation}
\begin{equation}
k = \frac{h_0}{(1-h)^2} \left\{1 + \frac{1-h_0}{h_0^2(1+h_0)} \left[2 \ln \left(\frac{1-h_0}{1-h}\right)
+ 2(h_0-h) + \frac{h_0^2 - h^2}{2}\right] \right\}, \label{2cf3}
\end{equation}
\begin{eqnarray}
l &=& \frac{h_0^2\sqrt{1+h_0}}{(h_0-1)^{\frac{3}{2}-\frac{(1-h_0)^2}{2h_0^2(1+h_0)^2}}} \exp \left[\frac{(1-h_0)^2(4+h_0)}{8h_0(1+h_0)^2}\right] \nonumber \\
&\times& \frac{1}{(1-h)^2}\left\lgroup \left\{1 + \frac{1-h_0}{h_0^2(1+h_0)} \left[2 \ln \left(\frac{1-h_0}{1-h}\right) + 2(h_0-h) + \frac{h_0^2 - h^2}{2} \right] \right\}^2 - \frac{h^2}{h_0^2}\right\rgroup, \label{2cf4}
\end{eqnarray}
\begin{equation}
\rho = - \frac{2(1-h)^{3-\frac{(1-h_0)^2}{h_0^2(1+h_0)^2}}}{\kappa} \left[\frac{(1-h_0)^2}{2h_0^2(1+h_0)^2} + \frac{1-h}{h(1+h)^3}\right] \exp \left[-\frac{(1-h_0)^2}{4h_0^2(1+h_0)^2}h(4+h)\right], \label{2cf5}
\end{equation}
\begin{equation}
P_\phi = - \frac{2(1-h)^{3-\frac{(1-h_0)^2}{h_0^2(1+h_0)^2}}}{\kappa} \left[\frac{(1-h_0)^2h}{2h_0^2(1+h_0)^2} + \frac{1-h}{(1+h)^3}\right] \exp \left[-\frac{(1-h_0)^2}{4h_0^2(1+h_0)^2}h(4+h)\right], \label{2cf6}
\end{equation}
\begin{equation}
D=\frac{h}{(1-h)^2}=r+\frac{h_0}{(1-h_0)^2}, \label{2cf7}
\end{equation}
\begin{equation}
h=\frac{1+2\left[r+\frac{h_0}{(1-h_0)^2}\right]-\sqrt{1+4\left[r+\frac{h_0}{(1-h_0)^2}\right]}}{2\left[r+\frac{h_0}{(1-h_0)^2}\right]}, \label{2cf8}
\end{equation}
\begin{equation}
h'=\frac{1}{\left[r+\frac{h_0}{(1-h_0)^2}\right]^2}\left\{\frac{1+3\left[r+\frac{h_0}{(1-h_0)^2}\right]}{\sqrt{1+4\left[r+\frac{h_0}{(1-h_0)^2}\right]}} -1\right\}, \label{2cf9}
\end{equation}
\begin{equation}
r=\frac{h}{(1-h)^2}-\frac{h_0}{(1-h_0)^2}. \label{2cf10}
\end{equation}

The allowed ranges for the parameter $h_0$ and for the function $h(r)$ are
\begin{equation}
-1<h_0<-0.42, \label{2cf11}
\end{equation}
\begin{equation}
-1<h_{\Sigma}<h<h_0<0. \label{2cf12}
\end{equation}
Moreover, it is easy to verify that the weak energy density condition is satisfied by this class of solutions provided (\ref{2cwe4}) is verified by $h$ for each value of $h_0$ defining a given solution of the class.

\section{Class 1 versus Class 2--(ii) solutions} \label{versus}

These classes are distinguished through the assumption made for the expression of the $f$ metric function from which they are built according to the method displayed in Sec. \ref{gm}.

The solutions pertaining to both classes are determined by a unique parameter, $h_0$, the ratio of the non-zero component of the pressure over the energy density evaluated at the symmetry axis. As it has been stressed in Paper 2, this property is to be contrasted to the four independent parameters appearing in the vacuum Lewis solutions. This implies that the presence of a fluid with pressure reduces the number of degrees of freedom of the induced spacetimes. To any given value of $h_0$ corresponds a given solution and, in particular, a given energy density and a given pressure both evolving as functions of the radial coordinate whose expressions have been displayed here.

Another common property is the fact that an explicit expression involving the radial coordinate $r$ has been established for the $h(r)$ function in each class. This makes easier the calculations and analyses using these metrics and any quantities pertaining to these solutions.

As in Paper 2, the amplitude of the rotation scalar of the fluid on the symmetry axis $\omega_0$ is given by the integration constant $c$ which appears as a parameter in the vacuum exterior Lewis solution. This is consistent with the result of Appendix A of Paper 2 and, since $c$ and therefore $\omega_0$ are both displayed as functions of $h_0$, any of these quantities can be retained as a parameter defining each solution of the set.

Finally, both classes satisfy the axisymmetry and the regularity conditions, the weak energy condition and a proper matching to an exterior Lewis-Weyl vacuum. All these properties qualify them as describing well-behaved spacetimes, potentially able to be used for different physical applications. Even though the regularity condition can, in some instance, see, e. g., Paper 2 for examples, constitute an unnecessary constraint for otherwise well-behaved solutions, this is not the case here where it merely contributes to reducing the number of independent parameters.

Now, some other physical behaviours of the solutions are different depending on the class to which they belong.

For Class 1 fluids, the positive energy density decreases from the axis down to the outer boundary whose radial coordinate $r_{\Sigma}$ is either bounded by $r_1$ displayed here as a function of the $h_0$ parameter or is unbounded, depending on two different allowed intervals for $h$. Two subclasses can therefore be distinguished depending on the belonging of $h$, and therefore of $h_0$, to one or to the other interval. Here, the azimuthal component of the pressure is equally a decreasing function of $r$ while the strong energy condition, $P_\phi>0$ is strictly satisfied. Solutions exhibiting such features might possibly be useful for the study of the gravitational behaviour of cylinders of standard fluids. 

Conversely, for Class 2 fluids, the pressure is negative. Hence, their physical use is limited to less standard configurations, e. g., potentially, cosmic (super)strings. The ratio $h$ is decreasing from the axis to the boundary $\Sigma$, which means, since the pressure is negative, that its amplitude with respect to that of the energy density increases while departing from the axis. Since the expression for $\rho$ is a rather complicated function of $h_0$ and $h$ and since therefore no general behaviour can be induced independently of any actual value of $h_0$, determining such a behaviour is left to the study of each particular application when needed. Such a remark applies equally to the hydrodynamical quantities displayed in Sec. \ref{2chyd}.

\section{Other integration method and the example of the $h=const.$ class} \label{other}

The method for solving the field equation displayed in Sec. \ref{gm} does not cover the whole set of admissible solutions. Indeed, only solutions possessing  metric functions which can be written as analytical functions of the quantity $h(r)$ are involved. Solutions whose metric functions cannot be explicitly written as functions of $P_{\Phi}/\rho$ are excluded.

Now, in the present case of an azimuthally directed pressure, another set of solutions can be explored. Assuming a barotropic equation of state written as $h=h(r)$, instead of assuming an expression for the metric function $f(h)$, the forms of the Einstein and Bianchi equations allow the following reasoning. 

The addition of (\ref{G11}) and (\ref{G22}) yields (\ref{sol1}) which can be integrated by (\ref{sol3}). Then,  by inserting (\ref{sol3}) and $h(r)$ into the Bianchi identity (\ref{Bianchi3}), we obtain $f'/f$ as a function of $r$. Provided this expression is integrable, we obtain $\ln{f(r)}$ an then, by exponentiation, $f(r)$. The other quantities of interest are then obtained by an easy adaptation of the reasoning displayed in Sec. \ref{gm}.

As an example, the solutions generated by fluids with equations of state $h=const.$ are worked out in the following.

\subsection{Integration of the field equations with $h=const.$}

Starting from (\ref{sol3}), which we recall here as
\begin{equation}
D = r+c_2, \label{1}
\end{equation}
we can write
\begin{equation}
\frac{D'}{D} = \frac{1}{r+c_2}, \label{2}
\end{equation}
which we insert into (\ref{Bianchi3}) and obtain
\begin{equation}
\frac{f'}{f} = \frac{2h}{1+h}\frac{1}{r+c_2}, \label{3}
\end{equation}
which, since $h$ is here a constant, can be integrated to yield
\begin{equation}
f = c_f (r+c_2)^{\frac{2h}{1+h}}, \label{4}
\end{equation}
where $c_f$ is an integration constant.

Then we use our usual equation for $k$ written as
\begin{equation}
k = f \left(c_k - 2c\int_{r_1}^r \frac{D(v)}{f(v)^2} \textrm{d}v \right), \label{5}
\end{equation}
where we insert (\ref{1}) and (\ref{4}), and then integrate to obtain
\begin{equation}
k = c_f (r+c_2)^{\frac{2h}{1+h}}\left[c_k - \frac{c}{c_f^2} \frac{1+h}{1-h} (r+c_2)^{\frac{2(1-h)}{1+h}}\right]. \label{6}
\end{equation}

Now, $l$ proceeds from (\ref{sol14}) where we substitute (\ref{1}), (\ref{4}) and (\ref{6}) which gives
\begin{equation}
l = \frac{(r+c_2)^{\frac{2}{1+h}}}{c_f} - c_f (r+c_2)^{\frac{2h}{1+h}}\left[c_k - \frac{c}{c_f^2} \frac{1+h}{1-h} (r+c_2)^{\frac{2(1-h)}{1+h}}\right]^2. \label{7}
\end{equation}

Then, we insert (\ref{1}), (\ref{4}), (\ref{6}) (\ref{7}) and derivatives into (\ref{G11}) such as to obtain
\begin{equation}
\mu' = - \frac{2h}{(1+h)^2 (r+c_2)} - \frac{2 c^2}{c_f^2} (r+c_2)^{\frac{1-3h}{1+h}}, \label{8}
\end{equation}
which can be integrated, and the result exponentialized, which yields
\begin{equation}
\textrm{e}^{\mu} = \frac{c_{\mu}}{(r+c_2)^{\frac{2h}{(1+h)^2}}} \exp\left[- \frac{c^2}{c_f^2} \frac{(1+h)}{(1-h)} (r+c_2)^{\frac{2(1-h)}{1+h}}\right].  \label{9}
\end{equation}
By rescaling the $r$ and $z$ coordinates, we can reset $c_{\mu}$ to unity.

Now, to calculate the energy density $\rho$, we insert the above functions and their derivatives of interest into, e. g., (\ref{G00}) and obtain
\begin{equation}
\rho = \frac{4c^2}{\kappa c_f^2 (1+h)}\frac{1}{(r+c_2)^{\frac{2h(1+2h)}{(1+h)^2}}} \exp\left[\frac{c^2}{c_f^2} \frac{(1+h)}{(1-h)} (r+c_2)^{\frac{2(1-h)}{1+h}}\right].  \label{10}
\end{equation}
The pressure follows as
\begin{equation}
P_{\phi} = h \rho \label{11}
\end{equation}
At this stage, each solution corresponding to a given value of $h$ depends on four integration constants. This set will be reduced to one independent parameter through the implementation of the conditions already applied to the previous classes of solutions.

\subsection{Axisymmetry and regularity conditions}

The axisymmetry condition, $l \stackrel{0}{=}0$, applied to the above solution by setting $r=0$ in (\ref{7}) reads
\begin{equation}
\frac{c_2^{\frac{2}{1+h}}}{c_f}\left[1+2cc_k\frac{(1+h)}{(1-h)}\right] - c_k^2c_fc_2^{\frac{2h}{1+h}} + \frac{c^2}{c_f^3} \frac{(1+h)^2}{(1-h)^2} c_2^{\frac{2(2-h)}{1+h}}=0.  \label{12}
\end{equation}

The regularity condition given by (\ref{regc}) where we insert (\ref{7}) and (\ref{9}) taken for $r=0$ can be written as
\begin{eqnarray}
&& \frac{c_2^{\frac{2(1-h)}{1+h}}}{c_f^2(1+h)^2} - \frac{c_2^{\frac{2}{(1+h)^2}}}{c_f} + \frac{4c c_2^{\frac{2(1-h)}{1+h}}}{c_f^2(1+h)} \left[c_k - \frac{c}{c_f^2}\frac{1+h}{1-h}c_2^{\frac{2(1-h)}{1+h}}\right] + 2\left[\frac{2c^2}{c_f^2}c_2^{\frac{2(1-h)}{1+h}} - \frac{h}{(1+h)^2}\right] \left[c_k - \frac{c}{c_f^2}\frac{1+h}{1-h}c_2^{\frac{2(1-h)}{1+h}}\right]^2 \nonumber \\
&-& 4c \frac{h}{1+h} \left[c_k - \frac{c}{c_f^2}\frac{1+h}{1-h}c_2^{\frac{2(1-h)}{1+h}}\right]^3
+ c_f^2 \frac{h^2}{(1+h)^2}\frac{1}{c_2^{\frac{2(1-h)}{1+h}}}\left[c_k - \frac{c}{c_f^2}\frac{1+h}{1-h}c_2^{\frac{2(1-h)}{1+h}}\right]^4  \nonumber \\
&+& c_f c_2^{\frac{2h^2}{(1+h)^2}} \left[c_k - \frac{c}{c_f^2}\frac{1+h}{1-h}c_2^{\frac{2(1-h)}{1+h}}\right]^2 \exp\left[- \frac{c^2}{c_f^2} \frac{(1+h)}{(1-h)} c_2^{\frac{2(1-h)}{1+h}}\right] = 0.  \label{13}
\end{eqnarray}

\subsection{Hydrodynamical properties}

The nonzero component of the acceleration vector is given by (\ref{1c32}) where we insert (\ref{4}) and derivative such as to obtain
\begin{equation}
\dot{V}_1 = \frac{h}{(1+h)}\frac{1}{(r+c_2)}. \label{14}
\end{equation}
Its modulus follows as
\begin{equation}
\dot{V}^{\alpha}\dot{V}_{\alpha} = \frac{h^2}{(1+h)^2} \frac{1}{(r+c_2)^{\frac{2(1+h+h^2)}{(1+h)^2}}} \exp\left[\frac{c^2}{c_f^2} \frac{(1+h)}{(1-h)} (r+c_2)^{\frac{2(1-h)}{1+h}}\right]. \label{15}
\end{equation}

The square of the rotation scalar is given by (\ref{1c36}) where we insert (\ref{4}) and (\ref{9}) to obtain
\begin{equation}
\omega^2 = \frac{c^2}{c_f^2}\frac{1}{(r+c_2)^{\frac{2h(1+2h)}{(1+h)^2}}} \exp\left[\frac{c^2}{c_f^2} \frac{(1+h)}{(1-h)} (r+c_2)^{\frac{2(1-h)}{1+h}}\right]. \label{16}
\end{equation}
Now we apply the rotation scalar theorem as derived in Appendix A of Paper II, i. e., $\omega^2 \stackrel{0}{=} c^2$, which we write as
\begin{equation}
 c_f^2 c_2^{\frac{2h(1+2h)}{(1+h)^2}} = \exp\left[\frac{c^2}{c_f^2} \frac{(1+h)}{(1-h)} c_2^{\frac{2(1-h)}{1+h}}\right]. \label{17}
\end{equation}

As already recalled in Section \ref{c1hyd}, rigid rotation implies a vanishing shear.

\subsection{One parameter solutions}

For each value of $h$ determining a given equation of state, we have three constraint equations, (\ref{12}), (\ref{13}) and (\ref{17}), for four integration constants, $c$, $c_2$, $c_f$ and $c_k$. Depending on the considered problem, one can therefore keep any of these constants as the solution parameter, and express explicitly or implicitly the three other constants in terms of this parameter.

\subsection{Weak energy condition}

Imposing the weak energy condition, i. e, $\rho$ given by (\ref{10}) positive, implies
\begin{equation}
h > -1, \quad c_2 > 0. \label{18}
\end{equation}
The set of the appropriate equations of state is therefore restricted such as to include, in particular, all those with $h>0$ which are the most usually used. 

\subsection{Metric signature}

The metric function $\textrm{e}^{\mu}$, with $c_{\mu}= 1$ and the weak energy condition $c_2>0$ satisfied, is positive definite. Hence, for a well-behaved Lorentzian signature of the metric, such must be the other three metric functions.

The positiveness of the $f$ function implies $c_f>0$.

That of the $k$ function yields 
\begin{equation}
c_k > \frac{c}{c_f^2}\frac{(1+h)}{(1-h)}(r_{\Sigma} + c_2)^{\frac{2(1-h)}{1+h}}, \label{19}
\end{equation}
where $r_{\Sigma}$ is the radial coordinate of the bordering cylinder.

The positiveness of the $l$ function depends on the value of h and cannot be studied in a fully general setting. A possible analysis method runs as follows. Since the axisymmetry condition imposes $l \stackrel{0}{=}0$, the first derivative of $l$ with respect to $r$ must be positive for $r=0$ such as to imply $l$ increasing and therefore positive in the vicinity of the axis. Then, there are two possibilities. Either $l$ does not vanish anywhere else than at $r=0$, and thus $r_{\Sigma}$ can take any value at will. Or $l$ vanishes for some $r_{max} \neq 0$ and this imposes a maximum width to the fluid cylinder such that $r_{\Sigma} \leq r_{max}$.

\hfill

We have thus displayed and analyzed another class of exact solutions, for spacetimes sourced by the simplest case of barotropic equation of state. Notice that the existence of such a solution for $h=const.$ is specific to the azimuthal pressure case. In particular, it has been shown by C\'el\'erier and Santos \cite{CS20}, that an equation of state corresponding to $h=const.$ is ruled out in the case of an axially directed pressure. This is due to the specific form of the field equations, implying in addition an integrable Bianchi identity, that allows two ways to address the problem. The first is described in Sec. \ref{gm} and the second in the present section.

\section{Conclusions}\label{concl}

Following the investigations of the interior spacetimes sourced by stationary cylindrical anisotropic fluids initiated in \cite{D06,CS20} and in Papers 1 and 2, the rigidly rotating fluid case with the particular equation of state $P_r = P_z=0$ has been examined here. Two general methods for constructing such solutions to the equations of GR have been proposed. The introduction of a new auxiliary function $h(r)$, defined as the ratio of the pressure over the energy density, as a tool designed to facilitate the integration of the field equations is of great benefit. Together with that of the auxiliary function $D$, known for long and which can be found elsewhere in the literature, its use has allowed the general methods described here to be finalized.

To exemplify the first method, where a given expression for the metric function $f$ is assumed, two such classes of exact solutions have been exhibited under the form of functions of $h(r)$ for the metric, the density, the pressure and other physical quantities of interest. Explicit expressions for $h$ as a function of $r$ have also been displayed and can be used to obtain exact expressions for the metric and the physical quantities as functions of the radial coordinate $r$. To ease the comparison between this set of solutions and the ones displayed in the other articles of the series, the expressions of interest have been presented here as functions of $h$. Of course, as usual, these solutions are valid in a given system of coordinates which, however, has been chosen such as to allow a direct physical interpretation. The axisymmetry and regularity conditions on the axis have been examined and discussed. As for the axial pressure case \cite{C21,C22a}, the solutions match trivially to the Weyl class of the Lewis vacuum solution on a cylindrical hypersurface $\Sigma$ acting as a boundary for the fluid. This is important in view of potential further uses for astrophysical purposes.

A number of physical and mathematical properties of these solutions have been calculated for each class, among them the hydrodynamical quantities. According to a result displayed in Appendix A of Paper 2, the intermediate parameter $c$, expressed as a function of $h_0$, has been physically interpreted as the amplitude of the vorticity of the fluid on the axis.

Constraints issued from thorough analyses of features of the solutions, e.g., the metric signature or the behaviour of the $h(r)$ function, have led to a sorting out among the different subclasses and subcases appearing in the course of the calculations. At the end of the day, only Class 1, with energy density and pressure decreasing from the axis to the boundary, and Class 2-- subclass (ii), whose solutions exhibit different physical behaviour depending on the value of the ratio $h$ at the axis, satisfy the whole set of conditions. Indeed, the value $h_0$ of this ratio on the axis emerges as the only independent parameter by which any given solution in a class is determined.

Owing to the particular form of the field equations corresponding to the present case of  azimuthally directed pressure, another integration method has been displayed. Here, instead of the function $f(h)$ in the first method, an equation of state $h(r)$ is assumed. This method has been applied to the case of fluids with equation of state $P_{\phi}=h\rho$, $h=const.$, for which another class of exact analytical solutions has been exhibited. As done for Class 1 and Class 2, the physical constraints and properties of these solutions have been thoroughly examined.

One might of course object that, at first sight, the proposed scheme could seem special from a standard physical point of view. Save perhaps in the case of purely axial pressure where some hints for direct astrophysical applications have been given, the rough equations of state considered in turn can be discussed. One could wonder, e. g., what about a fluid with purely azimuthal or axial pressure whose amplitude varies with the radial coordinate while the radial pressure remains null? The interest of the present approach is not to exhibit solutions to be considered at face value, but to provide a set of exact solutions from which quasi-anisotropy in quasi-cylindrical objects could be better understood. Of course, given the non-linearity of Einstein's field equations, the use of solutions of this kind for the study of a generalised anisotropic fluid is not simple, but one can suspect that such exact solutions might be used as, e.g., starting points for numerical or perturbative approaches. In such designs, each principal stress might be no more required to be in turn the only non vanishing component, but merely to dominate the other two. Therefore the above remark formulated as a question should no more be considered as a drawback. A possible application to cosmic topological defects has also been suggested in Paper 2. For a general discussion of the results presented in this series of articles and of their possible applications, the reader is referred to forthcoming Paper 5.

\acknowledgments

The author wishes to acknowledge interesting comments from an anonymous referee which helped improve the paper over an initial version.

\end{document}